\documentclass[12pt]{article}
\usepackage{jheppub}

\usepackage{mathtools}
\usepackage{amssymb, amsthm, amsbsy, bbm}
\usepackage{dsfont}

\usepackage[percent]{overpic}
\usepackage{xcolor}

\theoremstyle{remark}

\theoremstyle{definition}

\def\sl#1{\setbox0=\hbox{$#1$}           % set a box for #1 
   \dimen0=\wd0                                 % and get its size
   \setbox1=\hbox{/} \dimen1=\wd1               % get size of /
   \ifdim\dimen0>\dimen1                        % #1 is bigger
      \rlap{\hbox to \dimen0{\hfil/\hfil}}      % so center / in box
      #1                                        % and print #1
   \else                                        % / is bigger
      \rlap{\hbox to \dimen1{\hfil$#1$\hfil}}   % so center #1
      /                                         % and print /
   \fi}    
   
\makeatletter
\renewcommand\d[1]{\mspace{2mu}\mathrm{d}#1\@ifnextchar\d{\mspace{-3mu}}{}\mspace{4mu}}
\makeatother

\newcommand{\calB}{\mathcal{B}}

\newcommand{\calK}{\mathcal{K}}

\newcommand{\calO}{\mathcal{O}}

\renewcommand\d{\partial}
\newcommand\beq{\begin{equation}}
\newcommand\eeq{\end{equation}}
\newcommand\be{\begin{equation}}
\newcommand\ee{\end{equation}}

\newcommand\ks{\calK_g}

\newcommand\bb{\mathbb{B}}
\newcommand\kb{\calK_{\partial}}

\begin{document}
%%%%%%%%%%%%%%%%%%%%%%%%%%%%%%%%%%%%%%%%%%%%%%%%%%%%%%%%%%%%%%%%%%%%%%%%%%%%
%%%%%%%%%%%%%%%%%%%%%%%%%%%%%%%%%%%%%%%%%%%%%%%%%%%%%%%%%%%%%%%%%%%%%%%%%%%%

\title{Boundary Kinematic Space}

\author[a]{Andreas Karch, }
\author[b]{James Sully, }
\author[a,c]{Christoph F.~Uhlemann, }
\author[a,d]{Devin G.~E.~Walker}

\affiliation[a]{Department of Physics, University of Washington, Seattle, WA, 98195-1560, USA}
\affiliation[b]{Department of Physics, McGill University, Montr\'eal, QC, H3A 2T8, Canada}
\affiliation[c]{Mani L. Bhaumik Institute for Theoretical Physics, Department of Physics and Astronomy,
University of California, Los Angeles, CA 90095, USA}
\affiliation[d]{Department of Physics and Astronomy, Dartmouth College, Hanover, NH 03755, USA}

\preprint{\today}

\emailAdd{akarch@uw.edu}
\emailAdd{sullyj@physics.mcgill.ca}
\emailAdd{uhlemann@physics.ucla.edu}
\emailAdd{dgew@u.washington.edu}

\abstract{ %
We extend kinematic space to a simple scenario where the state is not fixed by conformal invariance: the vacuum of a conformal field theory with a boundary (bCFT). We identify the kinematic space associated with the boundary operator product expansion (bOPE) as a subspace of the full kinematic space. In addition, we establish representations of the corresponding bOPE blocks in a dual gravitational description.  We show how the new kinematic dictionary and the dynamical data in bOPE allows one to reconstruct the bulk geometry. This is evidence that kinematic space may be a useful construction for understanding bulk physics beyond just kinematics.
}

\maketitle

%%%%%%%%%%%%%%%%%%%%%%%%%%%%%%%%%%%%%%%%%%%%%%%%%%%%%%%%%%%
\section{Introduction}
%----------------------------------------------------------
One of the most appealing aspects of AdS/CFT, besides its use for pragmatic reasons to access strongly-coupled physics, is the geometrization of CFT structures in the dual gravitational description.
Indeed, attempts to make conformal symmetries manifest in geometric constructions predate the AdS/CFT correspondences: Dirac's use of ``conformal space'' for writing covariant field equations \cite{Dirac:1936fq} and the use of asymptotically-AdS geometries to construct curvature invariants which are covariant under Weyl transformations by Fefferman and Graham \cite{Fefferman:2007rka} are two prominent examples.
Whenever constructions in AdS/CFT can build on such geometric structures in conformal field theories, the dictionary between gravity and the CFT side takes a particularly natural form: Dirac's conformal space allows for a natural identification of the conformal boundary of AdS$_{d+1}$ with the conformal class of the CFT geometry, and the Fefferman-Graham coordinates and constructions are ubiquitous in AdS/CFT.

A more recent program to geometrize CFT structures, and in particular the operator product expansion (OPE), was started in \cite{Czech:2015qta,Czech:2016xec,deBoer:2016pqk,daCunha:2016crm}. In these papers the notion of kinematic space was introduced as a space of doubled dimension with a metric of balanced signature, on which the basic building blocks of OPEs admit a natural description as fields obeying wave equations.
This notion of kinematic space can be defined for arbitrary CFTs, whether they admit a holographic description or not, and provides an intuitive way to understand OPEs.
Moreover, this new geometric structure, once established, allows for an elegant formulation of the basic dictionary underlying the AdS/CFT correspondences: The OPE blocks of the CFT correspond to fields on kinematic space, which can in turn be reinterpreted as the space of geodesics or minimal surfaces in the AdS geometry of the dual gravitational description.
The OPE blocks themselves are then mapped to bulk operators smeared over these geodesics or minimal surfaces.  Note that, in this example, everything has been fixed by symmetry: the bulk AdS geometry, the auxiliary kinematic space geometry, the OPE blocks, and the geodesic operators.

Beyond conformal kinematics, we would like to know %
whether kinematic space is still a useful organizational principle for studying AdS/CFT.
More specifically, we would like to know how the kinematic space construction can be extended to describe states other than the vacuum.
A state ($\psi$) other than the vacuum is distinguished, in part, by CFT operators that have non-vanishing vacuum expectation values (vevs),
\begin{equation}
\langle \calO_i \rangle_\psi \neq 0 \; ,
\end{equation}
that determine the dual bulk geometry.
Perhaps the simplest such example is the vacuum of a CFT with a boundary (bCFT), where the $SO(2,d)$ symmetry has been broken down to the smaller group $SO(2,d-1)$.
With $y$ the coordinate transverse to the boundary, operators can take a vev
\begin{equation}
\langle \calO_i ({x},y) \rangle = \frac{A_i}{(2y)^{2\Delta_i}} \; ,
\end{equation}
that is fixed by symmetry up to a free parameter $A_i$. Thus, the bCFT vacuum is characterized by vevs that depend on dynamical data of the theory.
Correspondingly, when the bCFT has a gravitational dual, the metric is only constrained to take the form
\begin{equation}
ds^2 = e^{2 A(r)} ds^2_{AdS_d} + dr^2 \; ,
\label{eq:slicing-intro}
\end{equation}
where the function $A(r)$ is again determined by dynamical, rather than kinematic, data.  In this paper, we ask:
\begin{enumerate}
	\item Is there a corresponding notion of kinematic space for bCFTs?
	\item Does kinematic space help reconstruct the emergent gravitational physics for suitable bCFTs?
\end{enumerate}

The main structural novelty in CFTs with boundary is the existence of a boundary operator expansion (bOPE), which expresses an operator at a generic point of the CFT geometry in terms of a series of boundary-localized operators.
Our first main result is the definition of a kinematic space associated with this bOPE and the identification of its holographic representation
in the dual bulk geometry.
We then use the dynamical data of the bOPE and our dictionary for bOPE blocks to demonstrate how the bulk geometry can be reconstructed from the CFT data.
We explain how the bOPE provides the spectrum and asymptotic values of a radial mode expansion of a bulk field.
This is a well-known set of data for solving analogous inverse problems in asymptotically-flat space.  The solution of the asymptotically-AdS inverse problem is expected to follow a similar pattern, but we leave the details for future work.

We note that complementary aspects of bCFTs and kinematic space have been studied in \cite{Czech:2016nxc,Bhowmick:2017egz}. In contrast to our focus on the bOPE and a kinematic operator dictionary, these works discuss the kinematic space for bulk geodesics ending in the ambient space and connections to tensor networks and entanglement entropy.

The paper is organized as follows.  We start with a review of the relevant aspects of kinematic space and the kinematic dictionary for AdS/CFT in Sec.~\ref{sec:rev-kinematic} and discuss the structure of OPEs in bCFTs in Sec.~\ref{sec:rev-bCFT}.  In Sec.~\ref{sec:bCFT-dictionary} we extend the notion of kinematic space to CFTs with boundary and in particular to blocks in the bOPE,
and establish the holographic dual for these bOPE blocks.  Next, in Sec.~\ref{bulk-recon} we discuss how the bOPE organizes the data of the bulk geometry, and allows one to reconstruct the warp factor in the metric.
We conclude with a brief discussion of open questions in Sec.~\ref{discussion}.

%%%%%%%%%%%%%%%%%%%%%%%%%%%%%%%%%%%%%%%%%%%%%%%%%%%%%%%%%%%
\section{The kinematic dictionary}\label{sec:rev-kinematic}
%----------------------------------------------------------

In this section we review the connection between the OPE blocks, $\calB^k$, of a CFT in $d$ dimensions and geodesic operators in the holographic dual.

%----------------------------------------------------------
\subsection{A kinematic space for OPE blocks}\label{sec:rev-ope-blocks}
%----------------------------------------------------------

OPE blocks are eigenfunctions of the conformal Casimir operator \cite{Dolan:2003hv}.
The generators $L_{AB}$ (where $A$, $B$ run over the $d+2$ directions of the linear space on which the conformal group acts as $SO(2,d)$ matrices) of the conformal group can be written as differential operators acting on a single coordinate\footnote{In anticipation of the boundary we are about to introduce,
we use $M$, $N=0,\ldots,d-1$ to label the $d$ spacetime dimensions $X_M$ of the CFT. In the presence
of a boundary we will break up $X_M=(x_{\mu},y)$ with $\mu=0,\ldots,d-2$ labeling the $d-1$ directions along the defect and $y$ the transverse direction.} $X^M$.
The conformal group has a quadratic Casimir operator $L_{AB} L^{AB}$.
When acting on a primary operator ${\cal O}(X_1)$ of dimension $\Delta$ and spin $l$ the Casimir operator pulls out the eigenvalue
\be
L_{AB,1} L^{AB,1}\,  {\cal O}(X_1) = - \left [ \Delta (\Delta-d) + l (l+d-2) \right ] \, O(X_1) \equiv C_{\Delta,l} \, O(X_1).
\ee
Since the descendants of $O$ are in the same representation of the conformal group as the primary $O$ itself, they all have the same eigenvalue under the action of the Casimir operator.
Now the conformal block should correspond to the contribution of a primary of dimension $\Delta_k$ and spin $l_k$ and all its descendants appearing in the OPE of ${\cal O}_i(X_1)$ and ${\cal O}_j(X_2)$. As in~\cite{Czech:2015qta,Czech:2016xec} we restrict to the case where $\Delta_i=\Delta_j=\Delta$. The block should therefore be an eigenfunction of the quadratic Casimir
\beq
\label{lsquared}
L_{12}^2 = (L_{AB,1} + L_{AB,2}) (L^{AB,1} + L^{AB,2})
\eeq
with eigenvalue $C_{\Delta_k,l_k}$. In terms of the standard generators of the conformal group, $D$ (scale), $P_{M}$ (translation), $K_{M}$ (special conformal) and $M_{MN}$ (rotation \& boost) the quadratic Casimir reads
\beq
\label{casimir}
L_{AB} L^{AB} = - 2 D^2 - (K \cdot P + P \cdot K) + M^{MN} M_{MN}.
\eeq
Using the standard representation of these as differential operators acting on scalar fields
\begin{eqnarray}
D&=&ix^{M} \partial_{M}, \quad P_{M} = - i \partial_{M}, \quad K_{M} = i (2 x_{M} (x^{N} \partial_{N})
- x^2 \partial_{M}),\nonumber \\   \quad M_{M N} &=& -i (x_{M} \partial_{N} - x_{N} \partial_{M} )
\label{eq:inversion}
\end{eqnarray}
it is straightforward to evaluate the operator $L_{12}^2$ from \eqref{lsquared} to be given by
\beq
L_{12}^2 = 2 I^{MN} (x_1 -x_2)^2 \, \partial_{x_1^{M}} \partial_{x_2^{N}} \, ,
\label{eq:diff-op}
\eeq
where $I_{MN}$ is the standard inversion operator
\beq
I_{MN}(x_1-x_2) = \left ( \eta_{MN} - 2 \frac{(x_1-x_2)_{M} (x_1-x_2)_{N}}{(x_1-x_2)^2}
\right ) .
\eeq
The OPE block $\calB^k$ thus obeys a quadratic differential equation
\begin{equation}
\left(L_{12}^2  - c_{\Delta,l}\right) \calB^k(X_1,X_2) = 0 \, ,
\label{eq:def_eq}
\end{equation}
where $L_{12}$ is the differential operator in \eqref{eq:diff-op}.  

Equation~\eqref{eq:def_eq} can be viewed as an equation of motion for a free scalar field as follows:
We promote the pair of spacelike-separated points that specify the OPE block to be coordinates on an auxiliary  `kinematic space,' $\ks$.\footnote{We restrict our attention to the kinematic space for spacelike-separated points, $\ks$.
More general constructions are discussed in \cite{Czech:2016xec}.}
With a pair of spacetime coordinates, this space is $2d$-dimensional.
The metric structure on $\ks$ has signature $(d,d)$ and is uniquely determined by conformal invariance.
It takes the form
\beq
\label{metric}
ds^2 =I_{MN}(x_1-x_2)  \frac{d x_1^{M} dx_2^{N}}{|x_1-x_2|^2}.
\eeq
We can now reexamine the equation of motion, Eq. \eqref{eq:def_eq}.
The quadratic differential operator that appears is the Laplacian in $\ks$,
and the equation of motion is then that of a free field in $\ks$ with mass $m_k^2 = -c_{\Delta,l}$:
\begin{equation}
\left(\square_{\calK} + m_k^2 \right) \calB^k(X_1,X_2) = 0 \, .
\label{eq:ope-eom}
\end{equation}
Together with the boundary condition $\lim_{X_1 \rightarrow X_2} \calB^k \sim |X_{12}|^{\Delta_k} \calO_k$, and a set of constraints,\footnote{Kinematic space has signature $(d,d)$, and so the usual equation of motion is no longer hyperbolic, but ultra-hyperbolic. To set up a well-posed boundary value problem, one must also provide a set of constraints \cite{Czech:2016xec,deBoer:2016pqk}. This is another way of stating that the $2d$-dimensional kinematic space has more degrees of freedom and redundantly encodes the $d+1$-dimensional bulk geometry. The constraints are best-understood in $d=2$ where the symmetry algebra factorizes and the extra constraint equation is just the second quadratic Casimir.} this equation completely defines the OPE block.

%------------------------------------------------------------
\subsection{A kinematic space for geodesic operators}\label{sec:rev-geod-ops}
%------------------------------------------------------------

A pair of boundary spacelike-separated points also defines a unique minimal geodesic in the bulk, namely the geodesic that ends on these points.
Thus, the points of $\ks$ also label bulk geodesics.

This allows us to assign a natural set of \emph{bulk} operators to points in kinematic space:
For any local bulk operator $\phi$, and a geodesic $\gamma(X_1,X_2)$, we can consider the \emph{X-ray transform} of that operator
\begin{equation}
R\left[\phi\right] \left(X_1,X_2 \right)=\int_{\gamma(X_1,X_2)} \!\!\!\!\!\! \!\!\! ds\; \phi   \; .
\label{eq:xray-transform}
\end{equation}
To characterize the geodesic operator $R\left[\phi\right]$, we would like to understand what equation of motion in $\ks$ it obeys.
This can be accomplished by noting that there is a simple intertwining relation for the X-ray transform \cite{helgason1999,helgason2011,Czech:2016xec}:
\begin{equation}
\square_{\mathcal{K}}R\left[\phi\right]=-R\left[\square_{{\rm AdS}}\phi\right] \; .
\label{eq:intertwinement}
\end{equation}
Applied to the equation of motion for a free field in AdS, this gives
\begin{equation}
\left(\square_{{\rm AdS}}-m_k^{2}\right)\phi =0 \quad\implies\quad \left(\square_{\mathcal{K}}+m_{k}^{2}\right)R\left[\phi\right]=0 \; .
\label{eq:intertwining-eom}
\end{equation}
Together with the boundary condition $\lim_{X_1 \rightarrow X_2} R\left[\phi_k\right] \sim |X_{12}|^{\Delta_k} \calO_k$, which just uses the standard local holographic dictionary as the geodesic shrinks to the boundary,
and a set of constraints equivalent to those obeyed by the OPE blocks (here known in the mathematical literature as John's equations), this equation of motion determines our geodesic operator.

\paragraph{The kinematic dictionary}

As we reviewed, the OPE block $\calB^k$ and the geodesic operator $R\left[\phi_k\right]$ obey the same equation of motion, and have the same boundary conditions and constraints:
\begin{align}
\left(\square_{\mathcal{K}}+m_{k}^{2}\right)\mathcal{B}_{k}=0 & \qquad\left(\square_{\mathcal{K}}+m_{k}^{2}\right)R\left[\phi_{k}\right]=0\nonumber \\
\mathcal{B}_{k}\sim\left|x_{12}\right|^{\Delta_{k}}\mathcal{O}_{k} & \qquad R\left[\phi_k\right]\sim \left|x_{12}\right|^{\Delta_{k}}\mathcal{O}_{k} \; .
\label{eq:kinematic-dictionary}
\end{align}
Thus, we can conclude that $\calB^k$ and $R\left[\phi_{k}\right]$ become equivalent operators upon the usual identification of bulk and boundary Hilbert spaces \cite{Czech:2016xec}.

%%%%%%%%%%%%%%%%%%%%%%%%%%%%%%%%%%%%%%%%%%%%%%%%%%%%%%%%%%%
\section{Operator products in bCFT}\label{sec:rev-bCFT}
%----------------------------------------------------------

In this section we review the structure of operator products in CFTs with boundary as developed in general dimension in \cite{McAvity:1995zd}, following earlier developments in 1+1 dimensions which are nicely reviewed in \cite{Cardy:2004hm}.

What is a boundary CFT? Instead of considering a CFT on the $d$-dimensional plane, we instead consider a CFT on the upper-half plane with, hence, a planar boundary.
Doing so requires one to specify boundary conditions, possibly including a coupling of the original CFT to a $(d-1)$-dimensional theory restricted to the boundary.
By an appropriate choice of boundary conditions and boundary theory, one can preserve a residual $SO(2,d-1)$ subgroup of the conformal symmetry that leaves the boundary invariant.
Such `conformal boundary conditions' control the universal long-distance behavior of many physical systems with natural boundaries, such as quantum impurities (most notably in the context of the Kondo model as reviewed nicely in \cite{Affleck:1995ge}), quenches \cite{Calabrese:2006rx}, and many other examples of boundary critical phenomena \cite{Cardy:2004hm}, or open strings and D-branes \cite{Polchinski:1995mt}.

The residual $SO(2,d-1)$ symmetry of a bCFT preserves the full conformal invariance of the boundary theory.
Away from the boundary, however, we must adjust our typical restrictive assumptions of unbroken conformal invariance.
As mentioned in the introduction, non-trivial expectation values are now consistent with the residual symmetry. If the boundary is at $y=0$, then a scalar operator of dimension $\Delta_i$ can have a vev
\begin{equation}
\langle \calO_i ({x},y) \rangle = \frac{A_i}{(2y)^{2\Delta_i}} \; ,
\end{equation}
controlled by a free parameter $A_i$.
Likewise, a scalar two-point function now has the general form
\begin{equation}
\langle \calO_i ({x}_i,y_i) \calO_j ({x}_j,y_j) \rangle = \frac{g(\xi)}{(2y_i)^{2\Delta_i}(2y_j)^{2\Delta_j}} \, .
\end{equation}
where $g(\xi)$ is an arbitrary function of the invariant cross-ratio
\begin{equation}
\xi = \frac{(x_i - x_j)^2 + (y_i - y_j)^2 }{4 y_i y_j} \, .
\end{equation}
To make an intuitive connection to the kinematics of a typical CFT, one can imagine mirroring the upper-half plane and the operator insertions across the boundary.
Then the bCFT one-point function has the form of a two-point function in the mirrored CFT and the bCFT two-point function has the form of a 4-point function in a mirrored CFT. The `mirroring' of the bCFT should make it clear that, while the product of two operators has an interesting dynamical expansion in a CFT, a single operator will have an interesting dynamical expansion in a bCFT.

For the remaining discussion we will use the following terminology: as in \cite{Aharony:2003qf}, we refer to the
``bulk" of the CFT as the ambient space, so we can reserve the term bulk for the holographically dual spacetime geometry.\footnote{
That is, the CFT lives in a $d$ dimensional ambient space with a $d-1$ dimensional boundary.
The dual gravitational description lives in a $d+1$ dimensional bulk.}
To avoid confusion we will also refer to the kinematic space of Sec.~\ref{sec:rev-kinematic} as ambient kinematic space.
The main new feature is the expansion of ambient-space operators
in terms of boundary-localized operators, the bOPE \cite{McAvity:1995zd},
and does actually not involve an operator product at all.

%----------------------------------------------------------
\subsection{bOPE}\label{sec:rev-bOPE}

In the boundary operator product expansion (bOPE) \cite{McAvity:1995zd},
a single ambient space operator ${\cal O}_{i}(x,y)$ is expanded in terms of boundary localized operators $o_n$ as
\begin{equation}
{\cal O}_{i}(x,y) = \sum_n \frac{c_i^n}{(2y)^{\Delta_i - \Delta_n}}\left(1 + a_{n,1} y^2 \square_x + a_{n,2} y^4 \square_x^2 + \ldots \right) o_n(x)\,,
\label{eq:bope}
\end{equation}
where $\square_x$ is the d'Alembertian in the boundary directions.
We have grouped together an infinite sum over boundary descendants, given by powers of the d'Alembertian acting on a primary.
The explicit form of this sum can be found in \cite{McAvity:1995zd}, and is completely fixed by the $SO(2,d-1)$ boundary conformal symmetry.
The numerical coefficients of the primaries we label as $c_i^n$, and are called the bOPE coefficients.\footnote{The precise bOPE coefficients depend on a choice of normalization for both the boundary and ambient operators. We normalize boundary operators so that they have canonical two-point functions: $\langle o_n(x_1)o_n(x_2) \rangle = |x_1 -x_2|^{-2 \Delta_n}$. We normalize the ambient operators so that far from the boundary they are also canonically normalized: $\lim_{y\rightarrow \infty }\langle \calO_i(x_1,y)\calO_i(x_2,y)\rangle = |x_1 -x_2|^{-2 \Delta_i}$.}

Note that, were we to do this expansion about another ambient point, this would just be a simple Taylor expansion:
\begin{equation}
{\cal O}_{i}(x,y) = \sum_{n=0}^\infty \frac{1}{n!}(y-y_0)^n\partial_{y_0}^n {\cal O}_{i}(x,y_0) \, .
\end{equation}
The operator dimensions and coefficients that appear in the Taylor expansion are entirely fixed. But, when we take $\partial_{y_0}^n {\cal O}_{i}(x,y_0)$ to the boundary, the derivative operators may mix with other boundary operators and acquire anomalous dimensions.
Thus, the fact that we are expanding about a point on the boundary means that the operator dimensions and bOPE coefficients are not pre-determined by the geometry.

As boundary conformal invariance completely fixes the contribution of all the descendants in terms of those of the primary, we introduce the compact notation:
\begin{equation}
\frac{{\cal O}_{i}(x,y)}{(2y)^{\Delta_i}} = \sum_n c^n_i\bb_n(x,y)\,.
\label{eq:bblocks}
\end{equation}
The blocks $\bb_n(x,y)$ implicitly depend on the location of the boundary, so, unlike $\mathcal O_i$ itself, they are not localized at fixed $(x,y)$.
Much like the ambient space conformal blocks discussed in \cite{Czech:2016xec} are bilocal, they depend on $(x,y)$ as well as $(x,0)$, and are non-local operators.
This will be reflected in the bulk dual as well. %

\begin{figure}[t]
	\begin{center}
		\includegraphics[width=3.5in]{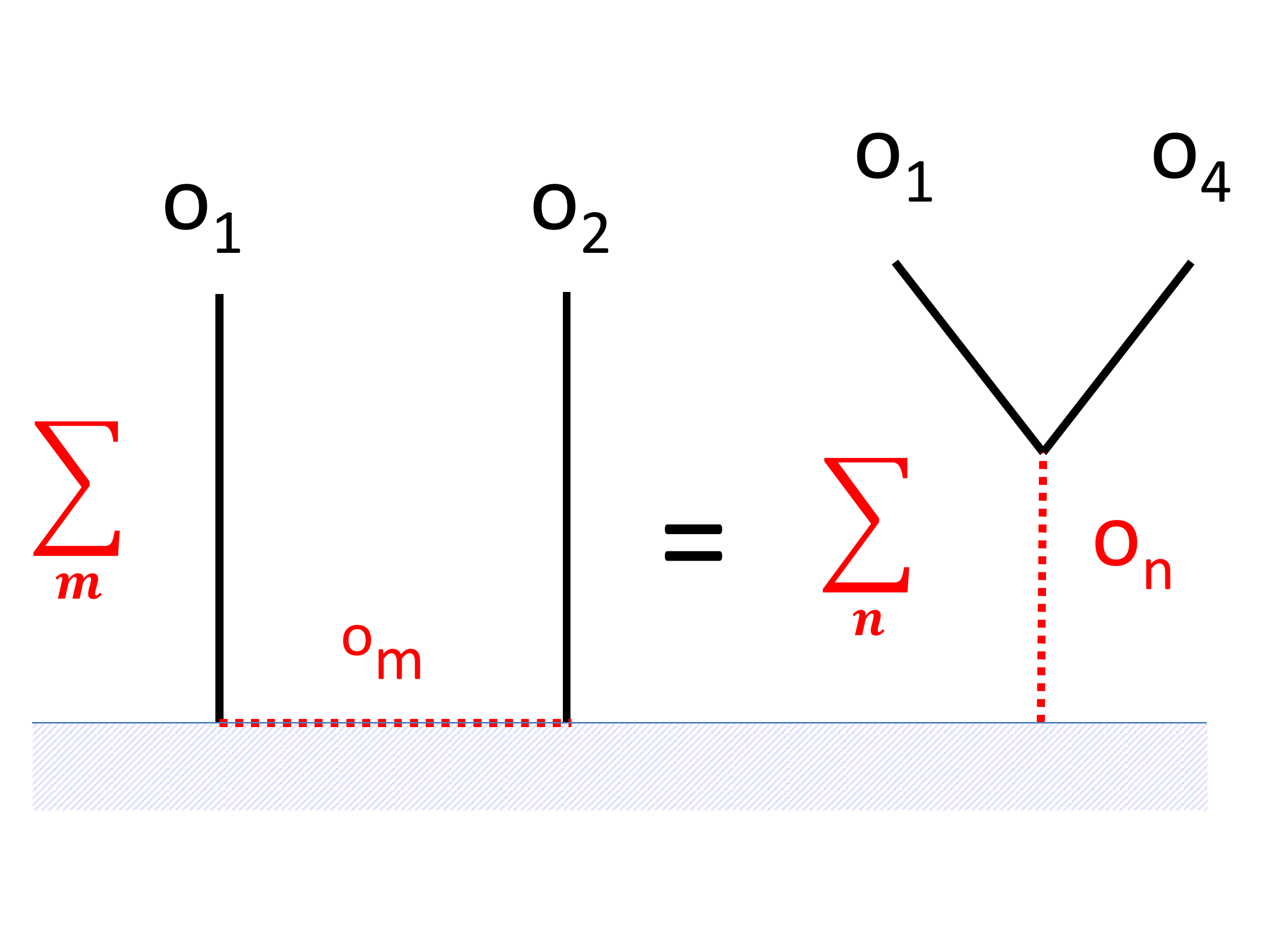}
	\end{center}
	\vskip -0.7cm \caption{Two-point function crossing symmetry underlying the conformal boundary bootstrap. In the presence of boundaries, one-point functions need not vanish, so the sum on the right hand side is non-trivial. Also, two-point functions can depend on one non-trivial conformally invariant cross-ratio and so there is already interesting information contained at that level.}
	\label{boundarybtrap}
\end{figure}

Standard conformal blocks are usually introduced as a contribution to the four point function rather than a non-local operator. One can think of the 4-pt function as being given by the 2-pt function of the non-local operators $\calB_k$.
This is one of the crucial ingredients in the conformal bootstrap program. Similarly, in conformal field theories with boundaries, the 2-pt function in the ambient space can be
expressed in terms of 2-pt functions of boundary operators.
As described in detail in \cite{Liendo:2012hy}, this allows a conformal bootstrap program for conformal field theories with boundaries already at the level of 2-pt functions.
The bootstrap equation, as depicted in fig.~\ref{boundarybtrap},
posits equivalence of expanding operators in terms of the standard OPE or the bOPE.

%------------------------------------------------------------
\subsection{Kinematic space for bCFT blocks}\label{sec:kinematic-bCFT}
%------------------------------------------------------------

The boundary conformal blocks defined in \eqref{eq:bblocks} obey a Casimir relation very similar to that of the standard conformal blocks. While the boundary blocks themselves had first been worked in \cite{McAvity:1995zd} by explicitly doing the sum of descendants, the Casimir method for boundary blocks was first introduced in \cite{Liendo:2012hy}.
Instead of the $SO(2,d)$ generators $L_{AB}$ one needs to consider the $SO(2,d-1)$ generators $L_{ab}$ that leave a planar defect invariant.
While the commutator algebra of this $SO(2,d-1)$ subgroup is the same as the conformal algebra of a $d-1$ dimensional field theory living on the defect, its action as differential operators acting on fields is distinct.
In particular, the dilation operator does not just rescale the $x_{\mu}$ coordinates along the defect, but also the $y$ direction orthogonal to it,
\beq
\label{op1}
\hat{D} = i (y \partial_y + x^{\mu} \partial_{\mu})\,,
\eeq
where we used hats to denote the generators of $SO(2,d-1)$ even though, as generators of a genuine subgroup, they are identical to their $SO(2,d)$ counterparts. Similarly we have
\beq
\hat{K}_{\mu} = i (2 x_{\mu} (X^{M} \partial_{N})
- X^2 \partial_{\mu})\,.
\eeq
Here $X^{M} \partial_{M} = y \partial_y + x^{\mu} \partial_{\mu}$ and $X^2 = x^{\mu} x_{\mu} + y^2$.
The remaining generators take their expected form
\beq
\label{op3}
 \hat{P}_{\mu} = - i \partial_{\mu},\quad \hat{M}_{\mu \nu} = -i (x_{\mu} \partial_{\nu} - x_{\nu} \partial_{\mu} )\,,
\eeq
and the quadratic Casimir of the boundary subalgebra is
\beq
L_{\partial}^2 = - 2 \hat{D}^2 - (\hat{K}^{\mu} \hat{P}_{\mu} + \hat{P}^{\mu} \hat{K}_{\mu}) + \hat{M}_{\mu \nu} \hat{M}^{\mu \nu} \; .
\label{eq:boundary-casimir}
\eeq

A boundary OPE block is a contribution to the expansion \eqref{eq:bblocks} of an ambient space operator in terms of defect operators that is an eigenfunction of the defect conformal Casimir, $L_{\partial}^2$, with eigenvalue
\beq
A_{\Delta} = \left . C_{\Delta,l=0} \right |_{d \rightarrow d-1} = - \Delta (\Delta - d +1).
\eeq
Spelling out $L_{\partial}^2$ acting on scalar blocks in terms of the differential operators of \eqref{op1}-\eqref{op3}, we find that the boundary conformal blocks are eigenfunctions of the operator
\beq
L_{\partial}^2 =2 y^2(\partial_y^2 + \partial_{\mu} \partial^{\mu}) - 2 (d-2) y \partial_y = 2 \left ( y^{d} \partial_y y^{2-d} \partial_y + y^2 \partial_{\mu} \partial^{\mu} \right ) \; .
\eeq
Similarly to the discussion in sec.~\ref{sec:rev-ope-blocks}, we can interpret this as a Laplace operator on an auxiliary boundary kinematic space, $\kb$.
This space is $d$-dimensional and naturally equipped with metric
\beq ds^2_{{\cal K}_{\partial}} = \frac{1}{2y^2} (dy^2 + dx^{\mu} dx_{\mu}). \label{ads} \eeq
That is, boundary kinematic space is simply AdS$_d$ with
its standard signature $(d-1,1)$.\footnote{This fact has basically been realized in a very different context already in \cite{Aharony:2003qf}, where it was shown that a field obeying a scalar wave equation in AdS$_d$ can exactly reproduce the blocks of the bOPE. We will come back to this construction when we discuss the holographic realization of boundary kinematic space.}
Then, the bOPE blocks are free fields in $\kb$ with mass $m_k^2 = -A_{\Delta}$:
\begin{equation}
\left(\square_{\calK_\partial} + m_k^2 \right) \bb^k(x,y) = 0 \, .
\label{eq:bound-ope-eom}
\end{equation}
Unlike the other kinematic spaces, this differential equation is hyperbolic and the boundary-value problem is well-posed without additional constraints.
With the boundary condition $\lim_{y \to 0} \bb^k \sim y^{\Delta_k} o_k(x)$ it completely defines the OPE block.

%------------------------------------------------------------
\subsection{Boundary kinematic space as a subspace of \texorpdfstring{${\cal K}_g$}{Kg}}\label{sec:subspace}

The relation of boundary kinematic space to the regular (ambient space) kinematic space can easily be understood. In the discussion of kinematic space, we were implicitly assuming that the conformal field theory lives on flat Minkowski space. In this case, the only conformally invariant boundaries we can introduce are planar defects. To understand the structure of correlation functions in the presence of a planar boundary, one can resort to the method of images.
That is, when we want to study correlation functions including an operator ${\cal O}(X)$ with $X^{M}=(y,x^{\mu})$ in a boundary conformal field theory, we can usually ensure that the boundary conditions on the planar defect are obeyed by including a mirror operator $\hat{O}(Rx)$ inserted at the
reflected point
\beq Rx = (-y,x^{\mu}) .\eeq
The details of the boundary conditions and boundary dynamics are included in the properties of $\hat{O}$. The boundary block expansion can be thought of as the standard OPE between ${\cal O}$ and its mirror operator. Note
that we usually want to think of boundaries as codimension one objects in space; that is, the orthogonal coordinate $y$ is spacelike, and hence so will be the separation between $x$ and $Rx$. This suggests that boundary kinematic
space should be a submanifold of ${\cal K}_g$, the kinematic space of spacelike separated points we reviewed above, and not its timelike counterpart. The submanifold is given by the $d$ embedding equations
\beq x_2 = R x_1. \label{embedding} \eeq
With this identification we find
\beq dx_1^{\mu}=dx_2^{\mu} =: dx^{\mu}, \quad dy_1 =- dy_2 =: dy, \quad
I_{\mu \nu}=\delta_{\mu \nu}, \quad I_{yy}=-1, \quad I_{\mu y}=0
\eeq
and indeed the induced metric on the subspace \eqref{embedding} of ${\cal K}_g$ (with its metric given in \eqref{metric}) becomes
the AdS$_d$ metric of \eqref{ads}.

%%%%%%%%%%%%%%%%%%%%%%%%%%%%%%%%%%%%%%%%%%%%%%%%%%%%%%%%%%%
\section{A kinematic dictionary for bCFT}\label{sec:bCFT-dictionary}
%----------------------------------------------------------

In \ref{sec:kinematic-bCFT} we studied the quadratic Casimir equations and constructed an appropriate auxiliary space in which bOPE blocks became free fields.
Now, we construct the corresponding dual geodesic operators and complete the dictionary.
We begin by reviewing the dual holographic geometry for a suitable bCFT.

%----------------------------------------------------------
\subsection{Holographic duals for bCFTs}\label{sec:bulk-dual}
%----------------------------------------------------------

The holographic dual description for conformal field theories with conformally invariant boundaries was given in \cite{Karch:2000gx,Takayanagi:2011zk}. The $SO(2,d-1)$ conformal symmetry preserved by the defect ensures that the $d+1$ dimensional bulk metric can be written as an $AdS_{d}$ slicing:
\begin{equation}
ds^2 = e^{2 A(r)} ds^2_{AdS_d} + dr^2.
\label{eq:slicing}
\end{equation}
AdS$_{d+1}$ itself, dual to a conformal field theory on all of Minkowski space, can be written in this form. In this case the warp factor is given by $e^A=\cosh{r/L}$. This exhibits two manifest asymptotic regions at
$r \rightarrow \pm \infty$. In the field theory these two asymptotic regions are conformal to two halves of the flat boundary spacetime. Without loss of generality we can identify them with the $y<0$ and $y>0$ regions of Minkowski space, where $y$ is one of the spatial coordinates. In addition, we can also reach the boundary of the bulk spacetime by approaching the boundary on the AdS$_d$ slice at fixed $r$. This boundary component maps to the $y=0$ plane separating the left and right half of spacetime.

A more general warp factor $e^{A(r)}$ that asymptotes to $\cosh(r/L)$ both at $r=\pm \infty$ corresponds to an interface CFT, with two potentially different CFTs on the left ($y<0$) and right ($y>0$) half of Minkowski space being connected by conformal invariance preserving boundary conditions on the $y=0$ interface. The two CFTs can in general be different and additional matter can be present on the interface. Examples of such interface CFTs are the Janus solution \cite{Bak:2003jk} or the D3/D5 system of \cite{Karch:2000gx,DeWolfe:2001pq}.
The former describes an interface between 3+1 dimensional ${\cal N}=4$ super Yang-Mills (SYM) theory with two different values of the coupling constant on the two sides. The latter describes the addition of a small number (much less than $N_c$, the number of colors) of interface localized matter multiplets into ${\cal N}=4$ SYM with $SU(N_c)$ gauge group.
Such interface CFTs can always be interpreted as boundary CFTs via the folding trick: we can simply redefine the $y$ coordinate for the CFT living on
the left half of space to $-y$, so that both CFTs live on the same $y>0$ half space. This way we rewrite the interface CFT as a boundary CFT with the special feature that in the ambient space the action describes two completely decoupled sectors with interactions between the two sectors confined to the interface. So in principle any such interface CFT (iCFT) can be seen as a special example of a CFT with boundary (bCFT).

For a genuine CFT with boundary, the metric \eqref{eq:slicing} only has one asymptotic region, say at large positive $r$, corresponding to half space. As usual, the warp factor should asymptotically approach the $\cosh(r/L)$ form in this region. The CFT boundary itself is dual to, as before, the asymptotic region that is reached by approaching the boundary of the AdS$_d$ slice. To avoid the appearance of the second asymptotic region, the bulk geometry has to terminate at some $r_*$. There are various options of how this can be implemented

\begin{enumerate}
\item The simplest option is to impose a discrete identification on AdS$_{d+1}$ under $r \rightarrow -r$. In the CFT this identifies the two halves of space, giving a particular realization of a CFT with boundary. The bulk geometry ends at the orbifold locus, $r_*=0$.

\item Spacetime can end in a hard wall, which can be implemented in a bottom up-way by a brane with tension. This is the proposal of \cite{Takayanagi:2011zk}. These spaces are solutions to consistent classical equations of motion. As they stand, these solutions do not typically follow from a string theory embedding and are hence at best seen as toy models for the complicated internal geometry. Correspondingly it is difficult to establish a particular dual CFT Lagrangian.
\item One finds a smooth 10d solution of, say, IIB supergravity, whose non-compact part takes the form \eqref{eq:slicing} with the full geometry smoothly ending at a finite $r_*$ by internal cycles shrinking. In this case the simple $d+1$ dimensional metric \eqref{eq:slicing} is not sufficient and the full 10d metric is needed. For the case of ${\cal N}=4$ super Yang-Mills on half space with supersymmetry preserving boundary
    conditions corresponding to NS5 or D5 branes, the full solution has been found in \cite{D'Hoker:2007xy,D'Hoker:2007xz} and analyzed in \cite{Aharony:2011yc}.  The geometry is AdS$_4\times$S$^2\times$S$^2$ warped over a 2d Riemann surface $\Sigma$ and the metric takes the form
\begin{equation}
ds^2 = f_4^2 (u,v) ds^2_{AdS_4} + f_A^2 (u,v) ds^2_{S^2_A} + f_B^2 (u,v) ds^2_{S^2_B} + d \Sigma^2~,
\label{eq:superjanus}
\end{equation}
where $u$ and $v$ are the coordinates on $\Sigma$.
\end{enumerate}

%------------------------------------------------------------
\subsection{Geodesic operators for bCFT}\label{sec:bCFT-geod-ops}
%------------------------------------------------------------

We now identify the geodesics and geodesic operators that should be associated to boundary kinematic spaces.

%------------------------------------------------------------
\subsubsection{Geodesics}\label{sec:geod-bope}

%------------------------------------------------------------

In analogy to the interpretation of ${\cal K}_g$ as the space of geodesics connecting two points, we can similarly think of boundary kinematic space as the space of geodesics connecting an ambient point $(y,x_{\mu})$ and its mirror $(-y,x_{\mu})$. This is exactly what happens in the case of an orbifold identification or an iCFT, case 1) in our list of possible holographic realizations, and one can again get the block from an X-ray transform.

For the case of a wall, either the hard wall of case 2) or the geometric wall of case 3), space really ends at $r_*$ and there is no obvious geometric realization of the mirror point. The way space ends will induce a boundary condition on the geodesic, so that one geodesic between $(y,x_{\mu})$ and the wall will be singled out. We want to argue here that there is a unique bulk geodesic associated to any ambient space point $(y,x_{\mu})$ that is consistent with the boundary conformal invariance.

Let us first exhibit the special geodesic in our 4 examples and postpone a look at its symmetry properties for later. The case of (folded) iCFTs as well as our examples 1 (orbifold) and 2 (Takayanagi's bottom-up construction) are based on the AdS$_d$ sliced metric of \eqref{eq:slicing}. Without loss of generality we can use translation invariance in the directions along the slice to set $x_{\mu}=0$ and so we are interested in geodesics emanating from the boundary point $(y_0,0)$. By rotation symmetry in the $x_{\mu}$ space we expect the geodesic to be given by $x_{\mu}(r)=0$ and so it can be completely parameterized by $y(r)$, where we chose the metric on the AdS$_d$ slice as
\beq
\label{localslice}
ds^2 = \frac{dy^2 + dx^{\mu} dx_{\mu}}{y^2}.
\eeq
The effective geodesic Lagrangian for $y(r)$ becomes
\beq
{\cal L} = \sqrt{1 + \frac{(y')^2 e^{2A} }{y^2}}~.
\eeq
While the generic solution to the geodesic equation is quite complicated, it is easy to see that the equation of motion
\beq
\partial_r \left ( \frac{e^{2A} y'}{y^2 \sqrt{1 + \frac{(y')^2 e^{2A} }{y^2}}} \right ) = - \frac{(y')^2 e^{2A}}{y^3 \sqrt{1 + \frac{(y')^2 e^{2A} }{y^2}}}
\eeq
allows for the simple solution
\beq y(r) = y_0 \eeq
for all $r$. It is similarly easy to see that $y(u,v)=y_0$ with $x_{\mu}=0$ identically is also a solution to the geodesic equation in the more general geometry of \eqref{eq:superjanus}

Out of all the geodesics that end in the point $(y_0,0)$ this geodesic is clearly special. It exist in all holographic duals to bCFTs, regardless of details.
It is also the closest bulk manifestation we have of a geodesic connecting the point and ``its mirror" in the cases where there is no second asymptotic region.
If we extend the geometry into the unphysical region, we end up exactly with the mirror point. We would like to argue that this special geodesic is also singled out by symmetry. This is easiest to see when we work with global AdS$_d$ coordinates on the slice, that is instead of \eqref{localslice} we use
\beq
- \cosh^2 \rho \, dt^2 + d\rho^2 + \sinh^2 \rho \,d \Omega_{d-2}^2.
\eeq
Choosing our boundary point to be $\rho=0$, the center of AdS, the special geodesic\footnote{This special geodesic was already used in \cite{Azeyanagi:2007qj} to calculate the entanglement entropy associated with iCFTs.} reads
\beq \rho(r) = 0 . \eeq
At the point $\rho=0$ the full $SO(d-1)$ rotation symmetry associated with the $\Omega_{d-2}$ factor in the metric is preserved. Any non-trivial $\rho$ dependence would lead to reduced symmetry. So at least for this special boundary point, the constant geodesic is singled out by enhanced symmetry. These rotations are part of the boundary conformal group we want to preserve.
But, since AdS$_d$ is a maximally symmetric and homogeneous space, all points are equivalent. That is by a boundary conformal transformation any point in AdS$_d$, which is conformally equivalent to half-Minkowski space, can be mapped to $\rho=0$ by an AdS$_d$ isometry. In the bottom-up models one would have expected that out of all the geodesics that emanate from a given boundary point one gets
picked by a boundary condition imposed at the ``wall" at $r=r_*$. Our claim is that unless the geodesic that gets picked is the special $y=y_0$ geodesic, the boundary conditions imposed on the CFT are inconsistent with the symmetries of a bCFT. The boundary breaks the conformal symmetry.

To connect back to kinematic space, our proposal is that from the holographic bulk point of view kinematic space should be viewed as the space of these geodesics with enhanced symmetry, so that there is a unique enhanced symmetry geodesic associated with each boundary point.

\subsubsection{bOPE geodesic operators}

We would like to use the geodesics constructed in the previous subsection to define the analog of an X-ray transform and the corresponding geodesic operators for the gravitational dual of bCFTs.
First note that there is no analog of  the constraints (John's equations) that were required in the original CFT kinematic space. In the bOPE expansion of a scalar ambient space operator only scalar boundary operators are allowed by Lorentz symmetry \cite{Liendo:2012hy}. So no additional restriction arises from asking the block to correspond to a scalar operator - this is automatically true. We also see that this time the dimension of boundary kinematic space, $d$, is {\it less} than the dimension of the $d+1$ dimensional bulk space time.
Furthermore, the bulk geometry is no longer uniquely fixed by symmetry.
The metric \eqref{eq:slicing} contains a free function $A(r)$ and we expect this function to influence the X-ray transform.

In fact, a precise connection between scalar fields in AdS$_{d+1}$ and conformal blocks appearing in the bOPE has been established previously in \cite{Aharony:2003qf} and we can easily recast the results obtained in there in terms of a modified X-ray transform. Any bulk scalar field $\phi(r,y,x_{\mu})$ obeying a wave equation with bulk mass $M$ on the $d+1$ dimensional geometry \eqref{eq:slicing} can be decomposed by a separation of variables ansatz
\beq\label{sov}
\phi(r,y,x_{\mu}) = \sum_n \psi_n(r) \bar\phi_n(y,x_{\mu}).
\eeq
Here the $\phi_n$ are scalar fields on the AdS$_d$ slice
\beq\label{eq:AdSd-modes}
\Box_{AdS_d} \bar\phi_n = m_n^2 \bar\phi_n
\eeq
with mass $m_n^2$, which together with $f_n$ are the eigenvalues and eigenfunctions of a 1d linear operator depending on the warp factor $A(r)$
\beq
\label{eq:modeequation}
\psi_n'' + d A' \psi_n' + e^{-2A} m_n^2 \psi_n - M^2 \psi_n =0.
\eeq
Primes denote derivatives with respect to $r$.
The mode functions $\psi_n$ are eigenfunctions of a Hermitian operator and so can be chosen to be orthonormal
\beq
\int dr \, e^{(d-2)A(r)} \psi_n \psi_m = \delta_{nm}.
\eeq
Here, and in what follows, the $r$-integral goes form $-\infty$ to $\infty$ in the case of a holographic dual to an iCFT and from $r_*$ to $\infty$ for the dual of a genuine bCFT. One interesting aspect of the separation of variables is that the coordinate $y$ on the slice exactly plays the role of the coordinate $y$ of the field theory living on the boundary. This way the scalar eigenfunctions on AdS$_d$ directly encode a non-local operator on the boundary. So we expect the scalar eigenfunctions $\phi_n(y,x_{\mu})$ to exactly be the conformal blocks of the boundary field theory.
This has in fact been established to be true in \cite{Aharony:2003qf}. The space on the slice in this sense {\it is} kinematic space. The dimension of the operator dual to $\bar\phi_n$ appearing in the bOPE dual is given by the usual relation
for AdS$_d$/CFT$_{d-1}$
\beq \Delta_n (\Delta_n - (d-1)) = m_n^2 \eeq
in terms of the eigenvalues $m_n^2$. The leading near boundary behavior of the mode functions encodes the coefficients with which the given primary appears in the bOPE. That is, the mode decomposition \eqref{sov} encodes much richer information than the scalar wave equation in kinematic space itself: in addition to the blocks we also get the bOPE coefficients.

We can use this construction to define a weighted X-ray transform associated with the holographic dual to bCFTs. What we want the X-ray transform to do is to map a scalar field on the bulk AdS$_{d+1}$ to a scalar field on boundary kinematic space ${\cal K}_{\partial}$, which is AdS$_{d}$. Clearly the map has to be one-to-many, just by counting dependent variables. We can define a family of weighted X-ray transforms
\beq R_n \phi(\gamma) = \int_{\gamma} ds \, \psi_n(x) \phi(x). \eeq
That is, we integrate the AdS$_{d+1}$ function $\phi(x)$ along the enhanced symmetry geodesic $\gamma$ (which is uniquely parametrized by a field theory point $(y,x_{\mu})$) weighted by the eigenfunctions $\psi_n$, which encode the details about the warp factor.
Given the mode decomposition \eqref{sov}, the orthogonality of the modes, and the fact that our enhanced symmetry geodesic was just given by $y=const.$, we can calculate
\beq \label{eq:weighted-R}
R_n \phi(\gamma) = \int dr \, \psi_n \phi(r,y,x_{\mu}) = \bar\phi_n(y_0,x_{\mu}).
\eeq
It is then immediate that the weighted X-Ray transform satisfies the same equation as the bOPE block:
\begin{equation}
(\Box_{AdS_d} - m_n^2) R_n \phi = 0 \; .
\end{equation}
In parallel with the discussion of ambient kinematic space and eq.~(\ref{eq:intertwinement}),
we can write this as an intertwining relation, by simply using $\Box_{\mathrm{AdS}_{d+1}} f=M^2f$, as
\begin{align}
 M^2 \Box_{\mathrm{AdS}_d}R_n[f](\gamma)&=
 m_n^2R_n[\Box_{\mathrm{AdS}_{d+1}}f](\gamma)~.
\end{align}

\paragraph{Boundary Conditions}
It remains now to check that the weighted X-ray transform and the bOPE block satisfy the same boundary conditions.
The bOPE block, in the limit $y\rightarrow 0$, becomes simply
\begin{equation}
\lim_{y\rightarrow 0} \mathbb{B}_n(x,y) = (2y)^{\Delta_n} o_n(x) + O(y^{\Delta_n + 2})
\end{equation}
for the primary operator $o_n(x)$. For the weighted X-ray transform, we showed above that it is just a canonically normalized field, $\bar{\phi}_n(x,y)$, on AdS$_{d}$. The limit $y \rightarrow 0$ is then just the standard limit as a bulk operator approaches the boundary \cite{McAvity:1995zd}, giving
\begin{align}
\lim_{y\rightarrow 0} R_n
\phi(x,y)& = a_{d-1,\Delta_n} y^{\Delta_n} o_n(x) + O(y^{\Delta_n + 2}) \;\; , \nonumber\\ a_{d-1,\Delta_n} &=\frac{1}{2 \Delta_n - (d-1)}\sqrt{\frac{\pi^{(d-1)/2} \Gamma(\Delta_n-\tfrac{d-1}{2})}{\left(2 \Delta_n - (d-1)\right)\Gamma(\Delta_n)}} .
\end{align}
Matching the two boundary conditions, we finally conclude
\begin{equation}
R_n
\phi(x,y) = b_n \bb_n(x,y) \; \; \; , \;\;\; b_n = 2^{-\Delta_n} a_{d-1,\Delta_n}\,.
\end{equation}
This establishes the bulk-to-boundary dictionary for bOPE blocks and weighted X-ray transforms.

%%%%%%%%%%%%%%%%%%%%%%%%%%%%%%%%%%%%%%%%%%%%%%%%%%%%%%%%%%%
\section{Bulk Reconstruction}\label{bulk-recon}
%----------------------------------------------------------

We have now generalized the holographic dictionary for geodesic operators that was proposed in \cite{Czech:2016xec} to the bOPE expansion.
In the prior work, much use was made of the symmetries of the problem; indeed, the bulk AdS geometry is a homogeneous space fixed by symmetry.
Here, however, our bulk geometry depends on dynamical data.
It is thus already interesting that we can find a simple correspondence between bOPE blocks and simple geodesic operators.

We would like to do more though: How do we invert this data to find local bulk fields and to determine the local bulk geometry? And, how is this encoded in the CFT dynamical data?

To invert our geodesic transform and obtain the local bulk field, we must write
\begin{equation}
\phi(x_\mu,w,r) = \sum_n \psi_n(r) \bar{\phi}_n(x_\mu,w) \; .
\end{equation}
We have already shown precisely that
\begin{equation}
R_n \phi (x_\mu,y) = \bar{\phi}_n(x_\mu,y) = b_n \mathbb{B}(x_\mu,y) \; ,
\end{equation}
so that the only missing information to reconstruct the bulk field is knowledge of the radial mode functions $\psi_n(r)$.
The radial mode functions are solutions of the linear equation \eqref{eq:modeequation}, which we recall is
\begin{equation}
\psi_n'' + d A' \psi_n' + e^{-2A} m_n^2 \psi_n - M^2 \psi_n =0 \; ,
\label{eq:radial-repeat}
\end{equation}
where the eigenvalues $m_n^2$ are the known Casimirs of the bOPE block.
The only unknown data in this equation is the function $A(r)$, the warp factor of the metric  \eqref{eq:slicing}.

A closely related reconstruction problem is to determine the bulk metric, \eqref{eq:slicing}, which we recall is
\begin{equation}
ds^2 = e^{2 A(r)} ds^2_{AdS_d} + dr^2 \; .
\label{eq:slicing-repeat}
\end{equation}
This metric is determined up to the warp factor $A(r)$, and this missing information is the same missing information needed to determine the mode functions.\footnote{If the field is not free, and there are non-trivial expectation values turned on, then the differential equation will depend more generally on a potential function. It is these potential functions, then, that we would be interested in directly reconstructing for different bulk fields, with the warp factor to be determined as a consequence of the ensemble. }

\emph{In effect, we are asking how to `hear the shape of a drum' by reconstructing the bulk geometry and local fields from specified knowledge about the eigenfunctions. }
We will give a strong indication that this is possible.

%%----------------------------------------------------------
\subsection{Bulk Data from the bOPE}
%%----------------------------------------------------------

To understand if the bOPE allows us to hear the shape of the bulk, let us collect the information that the CFT readily provides about the bulk radial eigenfunctions:

The first piece of data is obvious.
As we have already established, the radial eigenvalues are the known bOPE Casimirs $\lbrace m_n^2 \rbrace_{n=0}^\infty $.
The next place to look for useful data in the CFT is the set of bOPE coefficients, $c^n_i$. If we take the boundary limit of the bulk field to obtain the ambient space operator  $\calO_{d,i}$ we find a relation
\begin{equation}
\calO_{d,i}(x_\mu,y) = \frac{a_{d,i}^{-1}}{(2y)^{\Delta_i}} \sum_n C_n \bar{\phi}_n(x_\mu,y) \; ,
\end{equation}
where $C_n$ is the leading coefficient of the radial mode function in the limit $ r\rightarrow \infty$:
\begin{equation}
\psi_n (r) = C_n (e^{r})^{-\Delta_i} +  O\left((e^{r})^{-\Delta_i-2} \right)  \; .
\end{equation}
Substituting the normalization of our bulk modes in terms of the OPE blocks, we then find
\begin{equation}
\calO_{d,i}(x_\mu,y) = \frac{a_{d,i}^{-1}}{(2y)^\Delta_d} \sum_n C_n b_{n} \mathbb{B}_n(x_\mu,y) \, .
\end{equation}
Thus the bOPE coefficients $c^n_i$ determine the bulk data
\begin{equation}
c^n_i = C_n \, 2^{-\Delta_n} \frac{ a_{d-1,n}}{a_{d,i}} \, ,
\end{equation}
where the only unknown quantity is the leading coefficient, $C_n$, of $\psi_n(r)$.
It will be useful to phrase this data in a different way. Let us rescale the mode functions so that $C_n=1$. Then we have that the bOPE coefficients fix the norm of the radial mode functions:
\begin{equation}
\alpha_n := \int dr \, e^{(d-2)A(r)} |\psi_n|^2 = \left(  \frac{1}{c^n_i} 2^{-\Delta_n} \frac{ a_{d-1,n}}{a_{d,i}} \right)^2 \, .
\end{equation}

These two sets of discrete data, $\lbrace m_n^2 \rbrace_{n=0}^\infty $ and $\lbrace \alpha_n \rbrace_{n=0}^\infty $ are commonly chosen as sufficient data for solving inverse Sturm-Liouville problems.
We briefly review this procedure below.

%%----------------------------------------------------------
\subsection{The Inverse Sturm-Liouville Problem}
%%----------------------------------------------------------

We will describe the Gelfand-Levitan-Marchenko method for solving the inverse Sturm-Liouville problem \cite{MR17:489c,Marchenko:1986:SOA:21428,MR933088}.
In the interest of brevity, we will ignore many subtleties that will not be directly relevant to the more general point we wish to illustrate. In appendix \ref{app:sturm} we give a loose derivation of the the algorithm in a more familiar language to physicists.
See \cite{Marchenko:1986:SOA:21428,MR933088}, for example, for mathematical details.

Consider the Sturm-Liouville boundary value problem
\begin{equation}
-\varphi^{''}(r) + q(r) \varphi(r) = \lambda \varphi(r)
\end{equation}
on the real line with $q(r)$ the potential, $\lambda$ the spectral parameter.
Let us assume we do not know the potential function $q(r)$, except that it vanishes outside some finite interval.
On the other hand, let us assume that we do know the set of spectral parameters $\lambda_n = -m_n^2$ for the bound states in the potential and the norm of the associated eigenfunctions
\begin{equation}
\alpha_n = \int_0^\infty dr | \varphi_n(r)|^2 \, ,
\end{equation}
where the eigenfunctions have been normalized such that
\begin{equation}
\lim_{r\rightarrow \infty} \varphi_n(r) = e^{-m_n r} + \ldots \; .
\end{equation}
From this discrete data we can construct an auxiliary function
\begin{equation}
\tilde R_d(r) = 2 \sum_{n=0}^N \alpha_n^{-1} e^{-2 m_n r} \, .
\end{equation}
As the potential vanishes asymptotically, we also have a continuous spectrum of eigenstates for which we must also have knowledge of the reflection coefficient, $\tilde R_c(r)$, %
which we define in Appendix \ref{app:sturm}.
Then, we can set
\begin{equation}
\tilde R(r) = \tilde R_c(r) + \tilde R_d(r) \, ,
\end{equation}
and solve the integral equation
\begin{equation}
K(y,r) + \tilde R(y+r) + \int_0^\infty \tilde R(r+y+s)K(s,r)  ds = 0  \;\; ,\;\; y > 0  \; .
\end{equation}
The solution of this equation determines the potential via
\begin{equation}
q(r) = - \frac{d}{dr} K(0,r) \, .
\end{equation}
Thus, we have used the set of spectral parameters for the bound states in the potential and the norm of the associated eigenfunctions, along with the continuum of non-bound states, to solve the inverse problem.

What do we learn from this example? We have an analogous set of data (eigenvalues and norms) for the mode functions in an asymptotically-AdS geometry.
In our case, as in a finite interval or confining potential, there are no continuous modes.
It seems reasonable to propose, then, that the bOPE contains sufficient information to solve our inverse problem.
Unfortunately, we are not aware of a solution of our particular inverse Sturm-Liouville problem in the existing mathematical literature.
It would be nice to extend the Gelfand-Marchenko-Levitan method to asymptotically-AdS boundary conditions, but this problem is beyond the scope of this paper.

\section{Discussion and Outlook}\label{discussion}
%----------------------------------------------------------
We have extended the concept of kinematic space to conformal field theories with interfaces, defects or a boundary.
As an example of a theory where the vacuum state is not fixed by kinematics alone, this constitutes a first step towards generalizing kinematic space beyond pure kinematics.
The most intriguing new feature in this class of theories, as far as operator product expansions are concerned, is the bOPE, for which we have constructed a boundary kinematic space which is naturally embedded into the ambient kinematic space and on which bOPE blocks are scalar fields satisfying a Laplace equation.
As holographic duals of the bOPE blocks we have identified bulk operators smeared over a special geodesic which is anchored on the conformal boundary at the point where the operator is inserted, and connects it to the extension of the CFT boundary into the bulk.
Remarkably, detailed knowledge of the bOPE expansions appears to be enough to reconstruct the dual bulk geometry, and we have outlined the reconstruction procedure, which amounts to solving an inverse Sturm-Liouville problem.

\paragraph{Kinematics vs. dynamics} We take the natural way in which the concept of kinematic space extends to bCFTs as indication that it can indeed capture more than purely kinematic information.
The crucial step in elevating this from an expectation to a solid statement will be to understand how the mapping of ambient space OPE blocks to smeared bulk operators is affected by the breaking of the ambient space conformal symmetry.
The bulk geodesic anchored at two given boundary points is clearly affected by the choice of state in the CFT, as that affects the bulk geometry.
The OPE block, on the other hand, is defined and mapped to its bulk representation as an operator.
If kinematic space is to be useful beyond kinematics, the choice of state must be encoded in the mapping between OPE block and smeared bulk operator.
bCFTs provide a good starting point to address this question, due to the large amount of preserved symmetry:
the defect conformal group already provides a large amount of symmetry, and far away from the defect even the full conformal symmetry can be employed \cite{Liendo:2012hy}.
Other examples where this could be investigated are finite temperature states or the Coulomb branches of supersymmetric theories.

\paragraph{Background-independent operators} We established that the bOPE block is dual to a geodesic operator, regardless of the particular bulk warp factor.
Thus, our bOPE dictionary provides an example of a \emph{background-independent} operator (although perhaps a rudimentary example of such), in the sense that its definition does not refer to a particular background.
We did not need to know the bulk geometry to write down the bOPE block operator, and it has a simple, well-defined interpretation for any warp factor.
It would be even more interesting to determine background-independent CFT operators whose bulk interpretation does not depend on smearing with a radial mode-function.

\paragraph{Consistency of operator products} bCFTs, in fact, naturally produce an even richer hierarchy of kinematic spaces, with non-trivial consistency relations between them:
In direct analogy with the ambient space OPE one can define an OPE purely among boundary fields, and a corresponding kinematic space of dimension $2(d-1)$. It is tempting to argue that in the bulk this kinematic space is realized by geodesics that are confined to e.g.\ the hard wall of the models in \cite{Takayanagi:2011zk}.
But whatever the representation, consistency relations like the one illustrated in fig.~\ref{boundarybtrap} impose non-trivial constraints on the holographic dual representations and the involved geodesics, and thus good consistency checks.

\begin{figure}[tb]
\begin{center}
\includegraphics[width=0.95\linewidth]{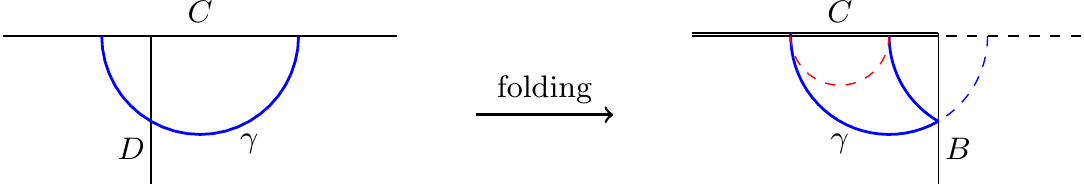}
\end{center}
\caption{
Folding of a geodesic across the defect in a dCFT to a long geodesic in a bCFT.
$C$ denotes the conformal boundary of the asymptotically-AdS geometry while $D$/$B$ labels the extension of the defect/boundary into the bulk.
}
\label{fig:long-geodesic}
\end{figure}

\paragraph{Long geodesics} Finally, topology also enters the stage.
As discussed in sec.~\ref{sec:bulk-dual}, a defect CFT can be mapped to a bCFT by folding, and the special geodesic relating to the bOPE can be understood as arising from the usual OPE of an operator with its mirror operator after folding.
Imagine now starting with operators asymmetrically inserted on both sides of the defect, as shown in fig.~\ref{fig:long-geodesic}.
After folding such a product is mapped to a product of ambient space operators, but the folding clearly instructs us to identify the corresponding OPE blocks not with the shortest geodesic connecting the insertion points, but with a ``long'' one going through the boundary.
This is reminiscent of the long geodesics playing a role in entanglement entropy calculations, and one may wonder if they play a role beyond the very special setup of a folded dCFT.
We leave those and other topics for the future.

\begin{appendix}
	
%%%%%%%%%%%%%%%%%%%%%%%%%%%%%%%%%%%%%%%%%%%%%%%%%%%%%%%%%%%%
\section{A derivation of the Gelfand-Marchenko-Levitan Method}\label{app:sturm}
%%----------------------------------------------------------

The brief description of the GLM method in the main text may seem rather opaque.
However, we can describe the method in the more familiar language of quantum mechanical scattering, following \cite{Marchenko:1986:SOA:21428,MR933088,terrytao}. We will demonstrate this in the simpler case where the potential has a continuous spectrum of scattering states, but no bound states.

Our differential equation has the form of a one-dimensional time-independent Schr\"o\-dinger equation on $\mathbf{R}$ with a potential $\hat{Q}$:
\begin{equation}
(\hat{H}_0 + \hat{Q})| \psi(k) \rangle   = k^2 | \psi(k) \rangle  \; ,
\end{equation}
where $\hat{H}_0$ is the free Hamiltonian. If the potential falls of sufficiently quickly, then this equation will have solutions that are purely incoming or outgoing asymptotically. Such solutions are found by using the Lipmann-Schwinger equation
\begin{equation}
| \psi^{(\pm)}(k) \rangle = | \pm k \rangle + \frac{1}{k^2 -\hat{H}_0} \hat{Q} |\psi^{(\pm)}(k) \rangle
\end{equation}
where $\hat{H}_0  | \pm k \rangle = k^2  | k \rangle$. We will choose the somewhat non-standard Green's function
\begin{equation}
G^{(\pm)}(r,r^{'}) = \Theta(\pm(r^{'}-r)) \frac{\sin \left(k(r-r^{'})\right)}{k}
\end{equation}
which is like a retarded/advanced propagator, except the roles of time and space have been exchanged. Using these propagators, the Lipmann-Schwinger equation becomes, in position space,
\begin{align}
\psi^{(+)}(k,r) = e^{ i k r} + k^{-1} \int_r^\infty dr^{'}\sin \left(k(r-r^{'})\right) Q(r^{'})  \psi^{(+)}(k,r^{'}) \\
\psi^{(-)}(k,r) = e^{- i k r} + k^{-1} \int_{-\infty}^{r} dr^{'}\sin \left(k(r-r^{'})\right) Q(r^{'})  \psi^{(-)}(k,r^{'}) \, .
\end{align}
(The solutions  to this integral equation are also known as the Jost solutions.)
If we consider a solution that scatters an incoming wave from $r = \infty$, so that it is purely outgoing at $r \rightarrow - \infty$, $\chi(k,r) \sim T(k) e^{-i k r}$, then we can write it in the form
\begin{equation}
T(k) \psi^{(-)}(k,r) = \psi^{(+)}(-k,r) + R(k) \psi^{(+)}(k,r) \; .
\label{eq:reflection}
\end{equation}
The equality necessarily holds because there are only two linearly-independent solutions. This equation defines the transmission and reflection coefficients $T(k)$ and $R(k)$.

We can now take the Fourier transform of \eqref{eq:reflection}, with respect to $k$, to find that the RHS becomes
\begin{equation}
\delta(t+r) + G(-t,r) + \check{R}(t-r) + \int_{-\infty}^\infty \check{R}(t-\tau)G(\tau,r) d\tau
\end{equation}
where
\begin{equation}
\check{R}(t) = \frac{1}{2 \pi}\int_{-\infty}^\infty e^{-ikt} R(k)  dk \; \; , \;\; G(t,r)= \frac{1}{2 \pi}\int_{-\infty}^\infty e^{-ikt} \psi^{(+)}(k,r) dk - \delta(t-r) \; .
\end{equation}
We can interpret $t$ quite literally as a time coordinate, and our Helmholtz equation becomes a wave equation.
At large $r$ where the potential vanishes, our solution is just $\delta(t+r) + R(t-r)$, which is just a left-moving delta-function wave and its reflection.
Thus, by causality, both our solution and $R(t-r)$ must vanish whenever $t<-r$. As a result, we find
\begin{equation}
G(-t,r) + \check{R}(t-r) + \int_{r}^\infty \check{R}(t-\tau)G(\tau,r) d\tau = 0 \;\;\; , \; \; t < -r
\end{equation}
We then use the transformation
\begin{equation}
G(t,r) = \frac{1}{2} K(\tfrac{1}{2}(t-r),r) \; \; , \;\; \tilde R(r) = 2\check{R}(-2r) \, .
\end{equation}
so that this equation becomes
\begin{equation}
K(y,r) + \tilde R(y+r) + \int_{0}^\infty \tilde R(r+y+ s)K(s,r) ds = 0 \; \; , \; \; y > 0  \; .
\end{equation}
This is the Gelfand-Levitan-Marchenko equation, allowing us to solve for $K(y,r)$ from the scattering data $R(k)$.
It remains then, only to check that
\begin{equation}
Q(r) = -\frac{d}{dr} K(0,r) \; ,
\end{equation}
which we can do by substituting $\check\psi^{(+)}(t,r)$ back into the wave equation that it solves.

\end{appendix}

\bibliographystyle{JHEP}
\bibliography{kinematic}

\end{document}